\documentclass[aps,prb,superscriptaddress,preprint]{revtex4-2}

\usepackage{amsmath,multirow}
\usepackage{graphicx}
\usepackage[figuresleft]{rotating}
\usepackage{bm,amsfonts,amssymb,makeidx}
\usepackage[T1]{fontenc}
\usepackage[inline]{enumitem}
\usepackage{chemformula}
\usepackage{amstext} 
\usepackage{soul}
\usepackage{array}   
\newcolumntype{C}{>{$}c<{$}} 
\newcolumntype{L}{>{$}l<{$}}
\def\ol#1{\overline{#1}{}}

\def\ti#1{\tilde{#1}{}}

\def\re{{\rm{e}}}
\def\ri{{\rm{i}}}

\def\ga{\alpha}
\def\gt{\theta}

\def\re{\mathrm{e}}\def\ri{\mathrm{i}}

\def\smatCJ#1{\begin{smallmatrix}#1\end{smallmatrix}}
\def\smatCD#1#2{\begin{smallmatrix}#1\\#2\end{smallmatrix}}
\def\smatCT#1#2#3{\begin{smallmatrix}#1\\ #2\\#3\end{smallmatrix}}
\def\smatCC#1#2#3#4{\begin{smallmatrix}#1\\ #2\\#3\\#4\end{smallmatrix}}

\def\ket#1{\mid~\!\!\!{#1}~\!\!\rangle}
\def\bra#1{\langle~\!\!{#1}~\!\!\!\mid}

\def\SO{\buildrel \rm SO \over \longrightarrow}

\begin{document}


\title{Fully linear band crossings at high symmetry points in layers: classification and role of spin-orbit coupling and time
reversal}


\author{N. Lazi\'c}
\email[]{natasal@ff.bg.ac.rs}
\affiliation{NanoLab, Faculty of Physics, University of Belgrade, Studentski trg 12, 11001 Belgrade, Serbia}
\author{V. Damljanovi\'c}
\email[]{damlja@ipb.ac.rs}
\affiliation{Institute of Physics Belgrade, University of Belgrade, Pregrevica 118, 11080 Belgrade, Serbia}
\author{M. Damnjanovi\'c}
\affiliation{NanoLab, Faculty of Physics, University of Belgrade, Studentski trg 12, 11001 Belgrade, Serbia}
\affiliation{Serbian Academy of Sciences and Arts, Kneza Mihaila St. 35, 11000 Belgrade, Serbia}

\date{\today}

\begin{abstract}
All of 320 layer groups, distributed into 80 clusters -- single/double ordinary/gray groups -- are used to complete systematization of linear (in all directions) band crossings and corresponding effective Hamiltonians in high-symmetry Brillouin zone points of layered materials, refining and expanding in literature existing data. Two- and four-dimensional effective Hamiltonians are determined by the allowed (half)integer (co)representations of the same dimension in the crossing point and one- or two-dimensional generic allowed representations. The resulting dispersion types (having isotropic or anisotropic form) are: single cone (with double degenerate crossing point and non-degenerate branches, or 4-fold degenerate crossing point with double degenerate conical branches), poppy-flower (4-fold degenerate crossing point with two pairs of non-degenerate mutually rotated conical branches), and a fortune teller (with nodal lines). Transition to double group, enabling to include spin-orbit interaction, results in various scenarios at high symmetry points: gap closing, gap opening, cone preserving, cone splitting \emph{etc}. Analogously, analyzing ordinary to gray group transitions, the role of time reversal symmetry is clarified.
\end{abstract}


\maketitle

\section{Introduction}

Interplay between symmetry and topology of band structures is among the most attractive topics in contemporary condensed matter physics. Besides topological insulators (TIs), nodal semimetals take a notable role, being a material realization of relativistic Dirac, Weyl, and Majorana particles~\cite{FelserAnnualRev17,ReMo18,GaoAnnualRev2019}, or lead to the emergence of unconventional quasiparticles~\cite{WiKi16,Beyond16,Triple16,WaAl16}. Characterized by band crossings (touching) points (lines) at Fermi level, with energies dispersing linearly, they have various interesting properties: Dirac points represent the interphases between topologically different insulating phases, Weyl points lead to semimetals with chiral anomaly, Fermi arc surface states \emph{etc}. Protected by crystal symmetries~\cite{VVDDAbrBen71,Manjes12,MuWe12,JaKe15,MiSm16,KiBa17,PaYa17,WiBr18}, these crossing points are robust with respect to various symmetry-preserving perturbations. Energy of the crossing cannot be predicted by symmetry alone; particularly important are those on Fermi level: when placed at (special) high symmetry
points (HSPs) in Brillouin zone (BZ), the material is known as a symmetry-enforced semimetal.

Leaving accidental degeneracy aside, the band crossings are within group theory related to the multi-dimensional allowed irreducible representations (IRs) of underlying symmetries. Geometrical transformations are gathered into ordinary crystallographic groups. When time reversal (TR) symmetry (either pure for paramagnetic systems, or combined with spatial symmetries for anti/ferromagnets) is included, gray or black-and-white magnetic groups~\cite{DIMMOCK62,BRADLEY68} are obtained; these are represented by irreducible corepresentations (coIRs), which have, besides unitary, additional anti-unitary operators. When spin space is included (spinfull case) to consider spin-orbit (SO) interaction half-integer irreducible (co)representations of double groups are assigned to electron bands.

The raising interest in exploring bands topology, including its symmetry based aspects, points out the necessity to systematize numerous particular studies, and fill in existing gaps. In particular, layer groups have been intensively used to predict Dirac and beyond-Dirac topological
semimetals~\cite{AsaHo11,JaKe15,Ja16,Ja16Ad,CaCr16,Wang17,PaYa17,YoWi17,Ja17,DamljanovicJPC20,LuoPRB20}, but still there is no complete overview of such symmetry-enforced band structures of layered materials. This thorough and systematic presentation will facilitate both numerical or experimental search for the materials with preferred symmetry and desirable band topology.

In this paper all band crossings with dispersion equations linear in all BZ directions around HSPs in quasi-2D crystals are singled out, with the corresponding effective low-energy Bloch Hamiltonians. We utilize allowed (co)IRs (calculated by POLSym code~\cite{Pol15}, and recently made available online~\cite{DGSITE}) of the symmetry groups of HSPs obtained by the action of layer (LGs), double layer (DLGs), and corresponding gray magnetic groups (gray LGs and gray DLGs) in BZ. It turns out that possible dimensions of (co)IRs, and therefore of the effective Hamiltonian models, are 1, 2, and 4. Among them, 2-dimensional ones may correspond to the Hamiltonians with completely linear band crossings hosting non-degenerate conical dispersion (1DC), while 4-dimensional (co)IRs support 2-degenerate conical (2DC), poppy flower (PF), or fortune teller (FT) shape of energy. The conical and PF dispersions may characterize semimetals, while the presence of FT indicates nodal line metal (where equienergetic lines cross in HSP). The relations between single and double groups (ordinary and gray) are described in order to facilitate studies of influence of spin orbit interaction (TR symmetry) to band topology. Results on the band crossings patterns (groups, HSPs, types, effects of spin and TR) are tabulated and discussed, stressing out the cases complementing or correcting those in literature.

The paper is structured as follows. Establishing basic concepts and notation, Sec.~\ref{SSymBloch} is a brief review of the group-theoretical apparatus within $k\cdot p$ theory. Then in Sec.~\ref{SLinDisp} we single out relevant high symmetry points in BZ of quasi-2D crystals, model Hamiltonians and the corresponding linear dispersions related to HSPs. Besides this, the robustness of the crossings (whether they are essential or not) is addressed in Sec.~\ref{SSOTR}, analyzing impact of spin and TR.

\section{Symmetry of effective Bloch Hamiltonian}\label{SSymBloch}

Following standard approach, we consider single-particle Hamiltonian $H$ invariant under symmetry group $G$ being one of the four types: ordinary group $G=L$, without TR symmetry $\theta$, is either purely geometrical layer group, or its double extension (to include spin space and SO interaction), while with $\theta$ it becomes (single or double) gray layer group $G=L+\theta L$ (for nonmagnetic systems).

On momentum $k$ from BZ $L$ acts by isogonal point group $P_I$. In this way LG makes stratification of BZ, singling out generic stratum, and special lines and points, each of them being fixed by characteristic little group (stabilizer) $L_k$ (a subgroup in $L$) of a representative momentum point $k$. Coset representatives $h$ from Lagrange partition $L=\bigcup_{h}h L_k$ generate star of $k$, and the set of representative points of all stars is irreducible domain (ID). Also, due to the trivial action of translations in BZ, all translations are in $L_{k}$, turning it into a (double) layer group.

While TR symmetry acts trivially on a position vector, it changes the sign of a momentum. Therefore, the addition of TR to the symmetry of layered systems in general changes stratification of the BZ, and three types of stabilizers $G_k$ (as subgroups in $G=L+\theta L$) may occur. For a TR invariant momentum (TRIM) $k$ the stabilizer is
\begin{enumerate*}[label=(\roman*)]
  \item gray group $G_k=L_k+\theta L_k$; otherwise, if $k$ is not TRIM, $G_k$ is either\label{Igray}
  \item black-and-white $G_k=L_k+\theta h L_k$ (if there is an nonidentity element $h$ such that $hk=-k$), or \label{Ibw}
  \item ordinary $G_k=L_{k}$ (either single or double) group.\label{Iord}
\end{enumerate*}
Notably, only in the latest case~\ref{Iord} the star is doubled due to the TR symmetry, while otherwise it remains the same (cases ~\ref{Igray} and ~\ref{Ibw}).

Commuting with the translational subgroup, the Hamiltonian reduces into the Bloch spaces. If $g$ belongs to $G_{k}$, meaning that $g$ stabilizes momentum $k$ up to the vector of inverse lattice, then
\begin{equation}\label{ECommBloch}
[D(g),H(k)]=0,
\end{equation}
where $D(G_k)$ is representation of stabilizer $G_k$ in the Bloch space and $H(k)$ is the Bloch Hamiltonian. The time reversal is antilinear operation in the state space, and therefore linear-antilinear representations $D(G_k)$ of magnetic little groups are considered: $D(L_k)=d(L_k)$ are linear operators, while the other elements are represented by antilinear operators $D(\gt hL_k)=d(\gt hL_k)\textsf{K}$, where $d(\gt h L_k)$ are linear factors and $\textsf{K}$ is the complex conjugation. Only matrix parts $d(L_k)$ and $d(\gt h)$ of all elements constitute co-representations. For gray groups, $h$ is the identity element, and $d^2(\gt)=\omega I$, where $\omega$ is $1$ for spinless, and $-1$ for spinfull cases ($I$ is identity matrix). Obviously, rewritten in the terms of co-representation for the antilinear coset the relation~\eqref{ECommBloch} is $d(g)H(k)d(g^{-1})=H^*(k)$ (for $g\in \theta hL_k$ ).

Consequently, symmetry provides that the corresponding Bloch Hamiltonian and the stabilizer representation are reduced in $|\ga|$-dimensional subspaces, where $|\ga|$ is the dimension of the allowed irreducible linear(-antilinear) representation $D^{(k,\ga)}(G_{k})$. Eigenvectors $\ket{k,\ga;a}$ of Bloch Hamiltonian:
\begin{equation}
H(k)\ket{k,\ga;a}=\varepsilon_\ga(k)\ket{k,\ga;a}, \nonumber
\end{equation}
are assigned by quantum numbers $(k,\ga)$ of linear(-antilinear) IRs (and allowed representations), meaning~\cite{JABOON} that:
\begin{equation}\label{ESAB}
D(g)\ket{k,\ga;a}=\sum_{a'}D^{(k,\ga)}_{a'a}(g)\ket{k,\ga;a'}. \nonumber
\end{equation}

Expansion of the Bloch Hamiltonian in the vicinity of HSP $k_0$ is
\begin{eqnarray}\label{ESeries}
H(k_0+k)&=&\sum_{n\geqslant 0}H^{(n)}(k_0+k),\\
H^{(n)}(k_0+k)&=&\tfrac{1}{n!}\sum_{p_1,\ldots,p_n}\tfrac{\partial^n H(k_0)}{\partial k_{p_1}\ldots \partial k_{p_n}}k_{p_1}\dots k_{p_n}, \nonumber
\end{eqnarray}
where $p_i=1,2$. Gathering terms with $n>0$ within perturbation $H'(k_0+k)$, an effective Hamiltonian is obtained with help of projector $P_\ga=\sum^{|\ga|}_{a=1}\ket{k_0,\ga;a}\bra{k_0,\ga;a}$ composed of the eigenvectors of unperturbed Hamiltonian $H^{(0)}(k_0)=H(k_0)$ (matrix with zero order term $n=0$). In the first perturbation order, the effective Hamiltonian is $H'_{\ga}(k)=P_\ga H'(k)P_\ga$, and the symmetry conditions~\eqref{ECommBloch} for each effective term $H^{(n)}_{\ga}$ of the expansion~\eqref{ESeries} becomes:
\begin{equation}\label{ECommBlochAllow}
D^{(k_0,\ga)}(g)H^{(n)}_{\ga}(k_0+k)D^{(k_0,\ga)}(g^{-1})=H^{(n)}_{\ga}(k_0+gk).
\end{equation}


As before, depending on the type of a considered system, $D^{({k}\ga)}=d^{({k}\ga)}$ is a unitary integer (spinless) or a half-integer (spinfull) IR of a (double) layer group, or, a linear-antilinear representation composed of the unitary matrix of coIR $d^{({k}\ga)}$ (multiplied by operator of complex-conjugation on the coset accompanied by TR) for a magnetic little group. In all these cases of layer groups the dimensions of IRs are 1, 2 or 4.

Stabilizer $G_{k_0+k}$ of a representative momentum $k_0+k$ from the generic stratum (dense in BZ) is a subgroup of $G_{k_0}$, and the subduced (co)representation obeys compatibility relations
\begin{equation}\label{ECompat}
d^{({k_0}\ga)}(G_{k_0})\downarrow G_{k_0+k}=\oplus_{i}f_i d^{({k_0+k},\ga_i)}(G_{k_0+k}),
\end{equation}
where $f_i$ is frequency number of the irreducible component $d^{({k_0+k},\ga_i)}$. For the elements of $G_{k_0+k}$ the symmetry condition~\eqref{ECommBlochAllow} becomes commutation causing that the energy branches in the vicinity of $k_0$ have degeneracies (1 or 2) of (co)representations $d^{({k_0+k},\ga_i)}(G_{k_0+k})$, while the degeneracy of the energy at crossing point $k_0$ coincides with the dimension (2 or 4) of (co)representation $d^{({k_0}\ga)}(G_{k_0})$. Finally, energies are invariants, which locally reads that $\varepsilon_{\ga_i}(k_0+k)=\varepsilon_{\ga_i}(k_0+g k)$ for $g \in G_{k_0}$.

\begin{figure}
\includegraphics[width=\columnwidth]{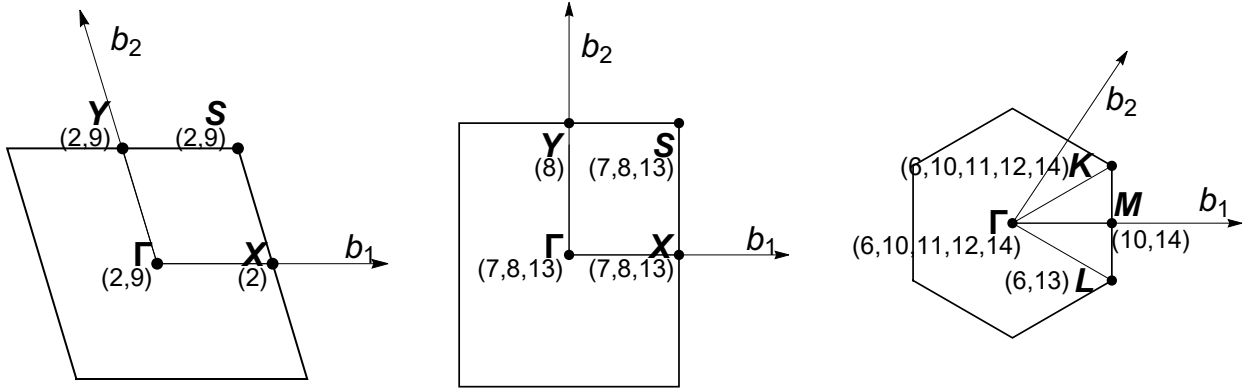}
\caption{\label{FStrata} High symmetry points. Each point is shown only in ID (equivalent copies from BZ are missing). Ordinals of IDs~\cite{DGSITE} are listed in brackets; IDs are associated to groups in Fig.~\ref{FSummary}. Left panel: oblique and c-centered rectangular (equal length of $b_1$ and $b_2$) groups; middle panel: rectangular-p and square (equal length of $b_1$ and $b_2$) groups; right panel: hexagonal groups.}
\end{figure}

\section{Linear dispersions}\label{SLinDisp}

\subsection{Effective Hamiltonian}\label{SSMethod}

At first, the forms of the effective Hamiltonians having completely linear dispersions in BZ around HSPs will be derived. According to~\eqref{ESeries}, non-zero parameters $w^p_{ab}=\tfrac{\partial [H_\ga(k_0)]_{ab}}{\partial k_{p}}$ allowed by symmetry condition~\eqref{ECommBlochAllow}, satisfying also hermiticity requirement $w^p_{ab}=w^{p*}_{ba}$, determine effective low-energy Hamiltonians linear ($n=1$) in momentum. Consequently, the matrix elements of such a Hamiltonian are $[H^{(1)}_{\ga}(k_0+k)]_{ab}=\sum_p w^p_{ab}k_p$; its linearly independent columns $w_{ab}=\left(w^1_{ab}, w^2_{ab}\right)^T$ define \emph{linearity rank}: number of BZ directions along which energies are linear in $k$. Obviously, completely linear dispersions have linearity rank 2. Since it is beyond the scope of the paper, herein the details about the dispersions are not studied, we only note in Sec.~\ref{SSOTR} which groups have linearity rank 1 (the vanishing linear term could be either nodal line when all higher order terms cancel, or of higher order dispersion). Linearity rank 0 refers to Hamiltonians without linear terms.

Instead of using absolute basis and parameters $w^p_{ab}$, it is more convenient to give the effective Hamiltonians in the basis of Hermitian matrices. With Pauli matrices $\sigma_i$ ($i=1,2,3$) and identity matrix $\sigma_0=I_2$, the effective 2D and 4D Hamiltonians (1D does not yield band crossing) are:
\begin{eqnarray}
H_2&=&\sum^3_{i=0}\sum^2_{p=1}v^p_{i}k_p\sigma_i,\label{EHam2D} \\
H_4&=&\sum^3_{i,j=0}\sum^2_{p=1}v^p_{ij}k_p(\sigma_i\otimes \sigma_j).\label{EHam4D}
\end{eqnarray}
Clearly, real parameters $v^p_i$ and $v^p_{ij}$ are bi-uniquely related to $w^p_{i}$ and $w^p_{ij}$, respectively. For each HSP and its allowed (co)IRs~\cite{DGSITE} of dimension 2 and 4, the symmetry allowed parameters $v^p_i$ and $v^p_{ij}$ are to be found. The task is performed assuming that layer is perpendicular to the $z$-axis. All calculations follow notation from Refs.~\cite{Pol15,DGSITE} (labels of HSPs, IDs), including (co)IRs.

\begin{sidewaysfigure}
\includegraphics[width=\textwidth,height=0.3\textheight]{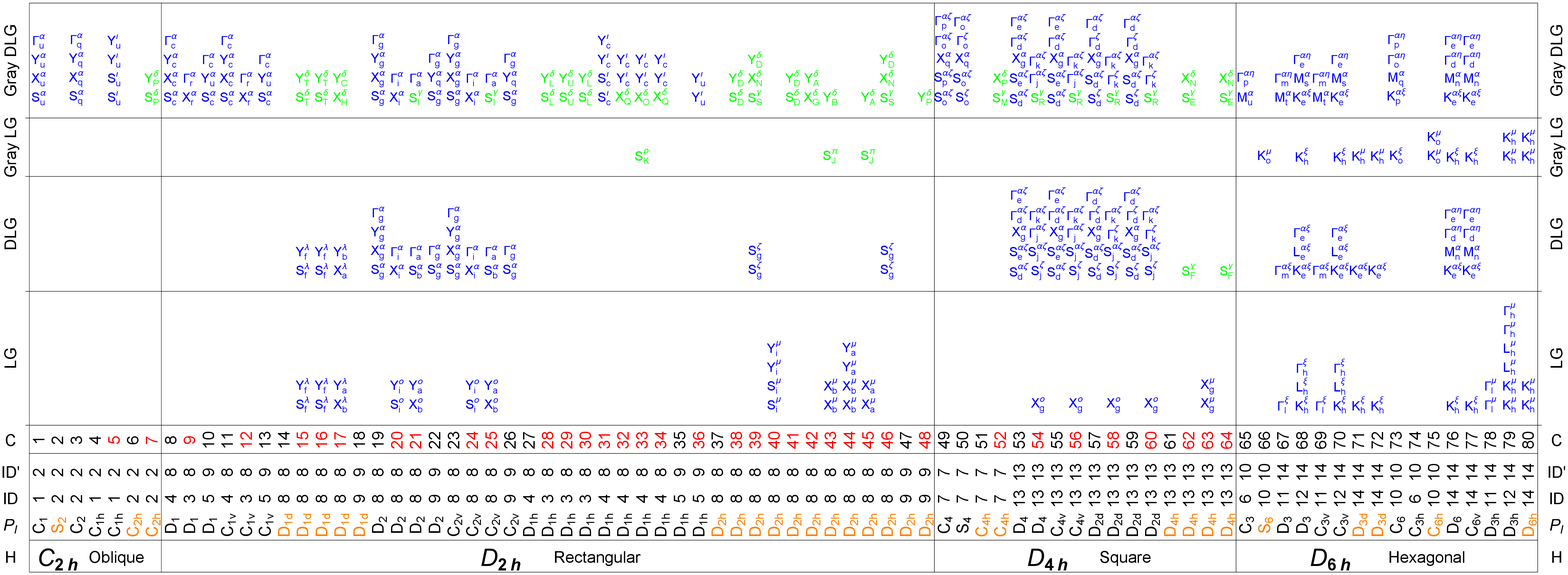}
\caption{\label{FSummary} HSPs (see Fig.~\ref{FStrata}) hosting completely linear dispersions obtained by action of LG, DLG, and their gray extensions in 2D BZ for each cluster C; ordinals are according to Ref.~\cite{VVDDITCE}. Dimensions of the allowed (co)IRs are distinguished by colours: blue stands for 2D, while green corresponds to 4D (co)IRs. The subscript is the label of the Hamiltonian model in Tables~\ref{THam2D} and~\ref{THam4D}. The superscripts correspond to the labels of SO transitions from the Tables~\ref{FSOOrdinary} and~\ref{FSOGray}. Also, the first column $H$ is holoedry (with lattice type) and isogonal groups $P_I$ are given in the second one. Those groups with inversion symmetry included are orange coloured, while non-symmorphic groups are singled out by red in column C. In the column ID and ID' are ordinals of irreducible domains of (gray) LGs according to the Ref.~\cite{DGSITE}.}
\end{sidewaysfigure}

Orthogonal part of any Euclidean transformation from arbitrary layer group leaves both the $xy$-plane and the $z$-axis invariant, having thus block-diagonal $2 \times2 + 1\times 1$ form. Action in 2D BZ is defined by the upper block. As a result, there is 10 different isogonal groups~\cite{DGSITE} composed of these $2\times2$ matrices, which, due to torus topology, yield 14 IDs of 2D BZ. IDs are the same for LG and DLG, while adding TR changes ID of noncentrosymmetric groups. Ordinals of the IDs are associated to the ordinary (gray) groups in the row ID (ID') in Fig.~\ref{FSummary}.

There are seven special points (Fig.~\ref{FStrata}): $\Gamma=(0,0)$, $X=(1/2,0)$, $Y=(0,1/2)$, $S=(1/2,1/2)$, $M=(1/2,0)$, $K=(1/3,1/3)$ and $L=(2/3,-1/3)$, with coordinates given in primitive basis $\{b_1,b_2\}$. They are distributed over 10 IDs: ID1 and ID3-ID5 have no HSPs; ID2 and ID8 have $\Gamma$, $X$, $Y$, and $S$; ID9 has $\Gamma$, $Y$, and $S$; ID7 and ID13 have $\Gamma$, $X$, and $S$; ID6 and ID12 have $\Gamma$, $K$, and $L$; ID10 and ID14 have $\Gamma$, $M$, and $K$; ID11 has $\Gamma$ and $K$. For oblique and rectangular-p groups with all non-symmorphic elements having fractional translations parallel to one direction (for group 45), $b_1$ is perpendicular to that direction (to the symmorphic reflection plane). For group 34 (32, 33, 43) $b_1$ is along axis (screw axis) of order two.

All of the HSPs are TRIM except $K$ and $L$. Therefore, the stabilizer of $\Gamma, X, Y, S, M$ is a gray group, while for $K$ it is either a black-and-white (in hexagonal gray LGs: 66, 67, 69, 71-73, 75-78, 80) or an ordinary group (in the hexagonal gray LGs: 65, 68, 70, 74, 79). Stabilizers of $L$ are ordinary, as it is the HSP only in ordinary groups (68, 70 and 79). In the most of the cases the stabilizer is the whole group, exceptions are points $Y$ in ID9, $K$ in ID11 and ID14, and $X$ in ID13 where it is a halving subgroup, and point $M$ in ID14 where the stabilizer is index-three subgroup.

\begin{table}\caption{\label{THam2D}Two-dimensional effective Hamiltonian forms. Non-vanishing symmetry adapted parameters $v^p_i$ in~\eqref{EHam2D} are defined in terms of independent constants $c_i$ (obtaining values in concrete problems). Symbol in column S is used in Fig.~\ref{FSummary} to identify model, while the number of the independent parameters, and corresponding dispersion equation are in collumns Par. and Eq; all energy branches are non-degenerate. Two coefficients $v^1_0$ and $v^2_0$, vanishing in all models, are omitted.}\begin{tabular}{|C|CCCCCC|CC|}
\hline
\text{S} & v_1^1 & v_2^1 & v_3^1 & v_1^2 & v_2^2 & v_3^2 & \text{Par.} & \text{Eq.} \\
\hline
 \text{a} & 0 & 0 & c_1 & 0 & c_2 & 0 & 2 & \eqref{EGen2DConeA} \\
 \text{b} & 0 & 0 & c_1 & c_2 & 0 & 0 & 2 & \eqref{EGen2DConeA} \\
 \text{c} & 0 & 0 & c_1 & c_2 & c_3 & 0 & 3 & \eqref{EGen2DConeA} \\
 \text{d} & 0 & -c_1 & 0 & c_1 & 0 & 0 & 1 & \eqref{EGen2DConeI} \\
 \text{e} & 0 & c_1 & 0 & c_1 & 0 & 0 & 1 & \eqref{EGen2DConeI} \\
 \text{f} & 0 & c_2 & 0 & 0 & 0 & c_1 & 2 & \eqref{EGen2DConeA} \\
 \text{g} & 0 & c_2 & 0 & c_1 & 0 & 0 & 2 & \eqref{EGen2DConeA} \\
 \text{h} & c_1 & 0 & 0 & 0 & c_1 & 0 & 1 & \eqref{EGen2DConeI} \\
 \text{i} & c_1 & 0 & 0 & 0 & c_2 & 0 & 2 & \eqref{EGen2DConeA} \\
 \text{j} & c_1 & -c_1 & 0 & -c_1 & -c_1 & 0 & 1 & \eqref{EGen2DConeI} \\
 \text{k} & c_1 & c_1 & 0 & -c_1 & c_1 & 0 & 1 & \eqref{EGen2DConeI} \\
 \text{l} & \sqrt{3} c_1 & -c_1 & 0 & c_1 & \sqrt{3} c_1 & 0 & 1 & \eqref{EGen2DConeI} \\
 \text{m} & c_1 & -\sqrt{3} c_1 & 0 & -\sqrt{3} c_1 & -c_1 & 0 & 1 & \eqref{EGen2DConeI} \\
 \text{n} & \sqrt{3} c_1 & -\sqrt{3} c_2 & 0 & c_1 & 3 c_2 & 0 & 2 & \eqref{EGen2DConeARot} \\
 \text{o} & c_1 & -c_2 & 0 & c_2 & c_1 & 0 & 2 & \eqref{EGen2DConeI} \\
 \text{p} & c_1 & c_2 & 0 & c_2 & -c_1 & 0 & 2 & \eqref{EGen2DConeI} \\
 \text{q} & c_1 & c_3 & 0 & c_2 & c_4 & 0 & 4 & \eqref{EGen2DConeARot} \\
 \text{r} & c_2 & c_3 & 0 & 0 & 0 & c_1 & 3 & \eqref{EGen2DConeA} \\
 \text{s} & \sqrt{3} c_2 & \sqrt{3} c_3 & \sqrt{3} c_1 & c_2 & c_3 & -3 c_1 & 3 & \eqref{EGen2DConeARot} \\
 \text{t} & \sqrt{3} c_2 & -\sqrt{3} c_3 & \sqrt{3} c_1 & -3 c_2 & 3 c_3 & c_1 & 3 & \eqref{EGen2DConeARot} \\
 \text{u} & c_3 & c_5 & c_1 & c_4 & c_6 & c_2 & 6 & \eqref{EGen2DConeARot} \\
 \hline
\end{tabular}\end{table}

\squeezetable\begin{table*}\caption{\label{THam4D}Four-dimensional effective Hamiltonian forms. Non-vanishing symmetry adapted parameters $v^p_{ij}$ in~\eqref{EHam4D} are defined in terms of independent constants $c_i$ (obtaining values in concrete problems). Symbol in column S is used in Fig.~\ref{FSummary} to identify model, while the number of the independent parameters, corresponding dispersion equation, and the degenracy of the branches are in collumns Par, Eq, and Deg. Ten coefficients $v^1_{00}$, $v^1_{01}$, $v^1_{03}$, $v^1_{32}$, $v^2_{00}$, $v^2_{01}$, $v^2_{03}$, $v^2_{12}$, $v^2_{22}$ and $v^2_{32}$, vanishing in all models, are omitted.}
\begin{tabular}{|C|CCCCCCCCCCCCCCCCCCCCCC|CCC|}
\hline
\text{S} & v_{02}^1 & v_{10}^1 & v_{11}^1 & v_{12}^1 & v_{13}^1 & v_{20}^1 & v_{21}^1 & v_{22}^1 & v_{23}^1 & v_{30}^1 & v_{31}^1 &
   v_{33}^1 & v_{02}^2 & v_{10}^2 & v_{11}^2 & v_{13}^2 & v_{20}^2 & v_{21}^2 & v_{23}^2 & v_{30}^2 & v_{31}^2 & v_{33}^2 & \text{Par.} & \text{Eq.} &\text{Deg.} \\
\hline
\text{A} & 0 & 0 & 0 & 0 & 0 & 0 & 0 & 0 & 0 & 0 & 0 & c_1 & 0 & 0 & 0 & c_2 & 0 & 0 & c_3 & 0 & 0 & 0 & 3 & \eqref{EGen2DConeA} &2 \\
\text{B} & 0 & 0 & 0 & 0 & 0 & 0 & 0 & 0 & 0 & 0 & 0 & c_1 & 0 & c_2 & 0 & 0 & c_3 & 0 & 0 & 0 & 0 & 0 & 3 & \eqref{EGen2DConeA} &2 \\
\text{C} & 0 & 0 & 0 & 0 & 0 & 0 & 0 & 0 & 0 & 0 & 0 & c_1 & 0 & c_3 & 0 & 0 & c_4 & 0 & 0 & 0 & c_2 & 0 & 4 & \eqref{EGen2DConeA} &2 \\
\text{D} & 0 & 0 & 0 & 0 & 0 & 0 & 0 & 0 & 0 & 0 & c_1 & 0 & 0 & c_2 & 0 & 0 & c_3 & 0 & 0 & 0 & 0 & 0 & 3 & \eqref{EGen2DConeA} &2 \\
\text{E} & 0 & 0 & 0 & 0 & 0 & -c_1 & 0 & 0 & 0 & 0 & 0 & 0 & 0 & c_1 & 0 & 0 & 0 & 0 & 0 & 0 & 0 & 0 & 1 & \eqref{EGen2DConeI} &2 \\
\text{F} & 0 & 0 & 0 & 0 & 0 & -\tfrac{c_1+c_2}{2} & 0 & 0 & \tfrac{c_1-c_2}{2} & 0 & 0 & 0 & 0 & \tfrac{c_1+c_2}{2} & 0 & \tfrac{c_1-c_2}{2} & 0 & 0 & 0 & 0 & 0 & 0 & 2 & \eqref{EGen4TwoCone} &1\\
\text{G} & 0 & 0 & 0 & 0 & c_2 & 0 & 0 & 0 & -c_3 & 0 & 0 & 0 & 0 & 0 & 0 & 0 & 0 & 0 & 0 & 0 & 0 & c_1 & 3 & \eqref{EGen2DConeA} &2 \\
\text{H} & 0 & 0 & 0 & 0 & c_2 & 0 & 0 & 0 & c_4 & 0 & 0 & c_1 & c_3 & 0 & 0 & 0 & 0 & 0 & 0 & 0 & 0 & 0 & 4 & \eqref{EGen2DConeA} &2 \\
\text{I} & 0 & 0 & 0 & 0 & c_3 & 0 & 0 & 0 & c_5 & 0 & 0 & c_1 & 0 & 0 & c_4 & 0 & 0 & c_6 & 0 & 0 & c_2 & 0 & 6 & \eqref{EGen4DPFA}&1 \\
\text{J} & 0 & 0 & 0 & c_3 & 0 & 0 & 0 & c_1 & 0 & 0 & 0 & 0 & c_2 & 0 & 0 & 0 & 0 & 0 & 0 & 0 & 0 & 0 & 3 & \eqref{EGen4DFTA} &1\\
\text{K} & 0 & 0 & 0 & c_4 & 0 & 0 & 0 & c_2 & 0 & c_1 & 0 & 0 & c_3 & 0 & 0 & 0 & 0 & 0 & 0 & 0 & 0 & 0 & 4 & \eqref{EGen4DFTA} &1\\
\text{L} & 0 & 0 & c_4 & 0 & 0 & 0 & c_6 & 0 & 0 & 0 & c_2 & 0 & 0 & c_3 & 0 & 0 & c_5 & 0 & 0 & c_1 & 0 & 0 & 6 & \eqref{EGen4DPFA} &1\\
\text{M} & 0 & c_1 & 0 & 0 & 0 & -c_2 & 0 & 0 & 0 & 0 & 0 & 0 & 0 & c_2 & 0 & 0 & c_1 & 0 & 0 & 0 & 0 & 0 & 2 & \eqref{EGen2DConeI} &2 \\
\text{N} & 0 & c_2 & 0 & 0 & 0 & c_3 & 0 & 0 & 0 & 0 & 0 & 0 & 0 & 0 & 0 & 0 & 0 & 0 & 0 & 0 & c_1 & 0 & 3 & \eqref{EGen2DConeA} &2 \\
\text{O} & 0 & c_2 & 0 & 0 & 0 & c_4 & 0 & 0 & 0 & c_1 & 0 & 0 & c_3 & 0 & 0 & 0 & 0 & 0 & 0 & 0 & 0 & 0 & 4 & \eqref{EGen4DFTA} &1\\
\text{P} & 0 & c_3 & 0 & 0 & 0 & c_5 & 0 & 0 & 0 & 0 & c_1 & 0 & 0 & c_4 & 0 & 0 & c_6 & 0 & 0 & 0 & c_2 & 0 & 6 & \eqref{EGen2DConeARot} &2\\
\text{Q} & 0 & c_3 & 0 & 0 & 0 & c_5 & 0 & 0 & 0 & c_1 & 0 & 0 & 0 & 0 & 0 & c_4 & 0 & 0 & c_6 & 0 & 0 & c_2 & 6 & \eqref{EGen4DPFA} &1 \\
\text{R} & c_1 & c_2 & 0 & 0 & c_3 & -c_3 & 0 & 0 & c_2 & 0 & c_1 & 0 & c_1 & c_2 & 0 & -c_3 & -c_3 & 0 & -c_2 & 0 & -c_1 & 0 & 3 & \eqref{EGen4DPFI} &1\\
\text{S} & c_2 & 0 & 0 & 0 & 0 & 0 & 0 & 0 & 0 & 0 & 0 & 0 & 0 & 0 & 0 & 0 & 0 & 0 & 0 & 0 & c_1 & 0 & 2 & \eqref{EGen2DConeA} &2 \\
\text{T} & c_3 & 0 & 0 & 0 & 0 & 0 & 0 & 0 & 0 & 0 & 0 & 0 & 0 & 0 & 0 & c_2 & 0 & 0 & c_4 & 0 & 0 & c_1 & 4 & \eqref{EGen2DConeA} &2 \\
\text{U} & c_3 & 0 & 0 & 0 & 0 & 0 & 0 & 0 & 0 & 0 & 0 & 0 & 0 & c_2 & 0 & 0 & c_4 & 0 & 0 & c_1 & 0 & 0 & 4 & \eqref{EGen4DFTA} &1\\
 \hline
\end{tabular}\end{table*}

Altogether we found 42 different effective Hamiltonians with completely linear dispersions at HSPs: 21 for 2D and 21 for 4D Hamiltonians are presented in Tables~\ref{THam2D} and~\ref{THam4D}. Number of nonzero coefficients $v^p_i$ and $v^p_{ij}$ may be 6, 4, 3, 2 or 1, as emphasized; the other vanish due to the symmetry. In particular, this includes those responsible for slope, which manifests that neither of the dispersions is tilted.

The results for all of 80 layer group clusters are summarized in Fig.~\ref{FSummary}. All HSPs hosting linearity rank 2 dispersions are listed, once for each of the associated allowed representations assigning/supporting such dispersions, with indicated effective model (subscript). The list of linear dispersions systematized in this way may be used for various analyses, and in the following sections some of them will be performed.

\subsection{Dispersion types}\label{SSRez}

Band crossings of the presented Hamiltonian models have linearity rank 2 with conical, poppy flower (both can be realized in isotropic or anisotropic forms) or fortune teller shape of dispersion. A conical dispersion corresponds to compatibility relations~\eqref{ECompat} with dimensions $2\rightarrow1 \oplus 1$ (1DC), or $4\rightarrow2\oplus2$ (2DC), while the both PF and FT are related to splitting dimensions $4\rightarrow1\oplus1\oplus1\oplus1$. The cases with 1DC and 2DC are usually referred to as Weyl and Dirac fermions respectively. PF consists of two mutually rotated non-degenerate anisotropic cones (some authors consider PF as generalized Dirac dispersion~\cite{CaCr16,JinPRL20}), while FT is composed of locally flat bands, with equi-energetic nodal lines.

For completeness, a brief overview of all types of dispersions (of linearity rank 2) are given, despite some of them have been already studied~\cite{JaKe15,Ja16,Ja16Ad,Ja17,DamljanovicJPC20}. For each model (row of the Tables~\ref{THam2D} and~\ref{THam4D}) the Hamiltonian matrix is formed according to~\eqref{EHam2D} or~\eqref{EHam4D} with non-vanishing $v^p_i$  and $v^p_{ij}$; it is expressed in terms of independent coefficients $c_1,\ldots,c_6$ (given in the row). As eigenvalues of these $k$-dependent matrices, the obtained dispersions are parameterized by coefficients $c_i$. For example $c_1( \sqrt{3}k_1+k_2)\sigma_1- c_2( \sqrt{3}k_1-3k_2)\sigma_2$ is the matrix for the two dimensional Hamiltonian in 14th row in Table ~\ref{THam2D} (symbol n).

\emph{General anisotropic 1DC} dispersion is
\begin{subequations}
\begin{equation}\label{EGen2DConeARot}
\varepsilon_{\pm}(k_1,k_2)=\pm \sqrt{a k^2_1+b k_1 k_2+c k^2_2},
\end{equation}
where $a, b, c$ are $c_i$-related parameters with ranges providing real energies. Equienergetic curves on this cone are ellipses with semi-axes $a'$ and $c'$ ($\tfrac{a}{\varepsilon^2}=\tfrac{\cos^2{\varphi}}{{a'}^2}+\tfrac{\sin^2{\varphi}}{{c'}^2}$, $\tfrac{c}{\varepsilon^2}=\tfrac{\sin^2{\varphi}}{{a'}^2}+\tfrac{\cos^2{\varphi}}{{c'}^2}$, $\tfrac{b}{\varepsilon^2}=2 \cos{\varphi}\sin{\varphi}\left( \tfrac{1}{{a'}^2}-\tfrac{1}{{c'}^2}\right)$), which are rotated with respect to the $k_1k_2$-coordinate system for the angle $\varphi$ between axes $a'$ and $k_1$. To illustrate, the Hamiltonian n (Table ~\ref{THam2D}) from the above example has dispersion~\eqref{EGen2DConeARot}, with $a=3(c^2_1+c^2_2)$, $b=\sqrt{12}(c^2_1-3c^2_2)$ and $c=c^2_1+9c^2_2$.

For $b=0$ the dispersion is still an anisotropic 1DC (but not rotated)
\begin{equation}\label{EGen2DConeA}
\varepsilon_{\pm}(k_1,k_2)=\pm \sqrt{a k^2_1+c k^2_2},
\end{equation}
which isohypses are ellipses with semi-axes $\tfrac{\varepsilon}{\sqrt{a}}$ and $\tfrac{\varepsilon}{\sqrt{c}}$. Finally, \emph{isotropic 1DC} is obtained by $a=c$:
\begin{equation}\label{EGen2DConeI}
\varepsilon_{\pm}(k_1,k_2)=\pm a |k|.
\end{equation}
\end{subequations}

As for 4D, \emph{general anisotropic PF} dispersion~\cite{DamljanovicJPC20} is:
\begin{subequations}
\begin{equation}\label{EGen4DPFA}
\varepsilon_{\pm,u}(k_1,k_2)=\pm \sqrt{a k^2_1+ u b |k_1 k_2|+c k^2_2},\quad u=\pm1.
\end{equation}
Substituting $a=c$ the \emph{isotropic PF} is obtained:
\begin{equation}\label{EGen4DPFI}
\varepsilon_{\pm,u}(k_1,k_2)=\pm \sqrt{a k^2+ u b |k_1 k_2|},\quad u=\pm1,
\end{equation}
while \eqref{EGen4DPFA} for $b^2=4ac$ becomes nodal line \emph{FT dispersion}~\cite{Ja17}:
\begin{equation}\label{EGen4DFTA}
\varepsilon_{\pm,u}(k_1,k_2)=\pm \left|\sqrt{a} |k_1|+ u\sqrt{c} |k_2|\right|,\quad u=\pm 1.
\end{equation}
Effective model Hamiltonian F from Tab.~\ref{THam4D} describes also isotropic PF but slightly modified:
\begin{equation}\label{EGen4TwoCone}
\varepsilon_{\pm,u}(k_1,k_2)=\pm \sqrt{\tfrac{(c^2_1+c^2_2) (k^2_1+k^2_2)+u |c^2_1-c^2_2||k^2_1-k^2_2|}{2}} ;
\end{equation}
substitution $k_1\pm k_2\rightarrow k_{\pm}$ reduces it to the form~\eqref{EGen4DPFI}. Here, positive $\varepsilon_{+,u}$ (as well as negative $\varepsilon_{-,u}$) branches are touched along the lines $k_1=\pm k_2$.

The rest of the 4D Hamiltonians result in 2DC (double degenerate cones described by equations discussed in 2D case).
\end{subequations}


\section{Analysis}\label{SSOTR}

Having at disposal all possible completely linear dispersions in the HSPs of layered systems, we analyse their interrelations. In this context the roles of SO coupling and TR symmetry are examined. In the group-theoretical language inclusion of spin can be seen as transition from single to double group, while TR relates ordinary and gray group.

\subsection{Spin-orbit interaction}\label{SSSO}

SO interaction is taken into account through the relation between integer and half-integer representations. Total space is tensor product of the orbital space with two-dimensional spin-half space, the later carrying spin representation $u(G_{k_0})\in\mathrm{SU}(2)$. Since composed of $\mathrm{SU}(2)$ matrices, $u$ can be either irreducible or reducible $u=u_1\oplus u_2$ ($u_i$ are irreducible). Hence, each integer irreducible (allowed) co-representation $d^{({k_0},\ga)}(G_{k_0})$ is multiplied by $u(G_{k_0})$, yielding a half-integer representation, either irreducible itself $d^{({k_0},\tilde{\ga})}(G_{k_0})$ (with frequency number $f^{\tilde{\ga}}=1$ in the decomposition below), or decomposed onto irreducible components (associated to ${k_0}$ and counted by $\tilde{\ga}$):
\begin{equation}\label{ESO}
d^{({k_0},\ga)}(G_{k_0})\otimes u(G_{k_0})=\oplus_{\tilde{\ga}} f^{\tilde{\ga}} d^{({k_0},\tilde{\ga})}(G_{k_0}).
\end{equation}


\begin{table*}\caption{\label{FSOOrdinary} Influence of SO coupling to the type of the splitting without TR symmetry: each row denotes a particular type (label in the column T is used as superscript in Fig.~\ref{FSummary}) of transition from spinless case (described in the next two columns by degeneracy $|\ga|$ at the crossing point, and linearity rank $L_{\ga}$) to the spinfull case (in the following columns: frequency number $f^{\tilde{\ga}}$ in decomposition~\eqref{ESO}, crossing point degeneracy $|\tilde{\ga}|$, and linearity rank $L_{\tilde{\ga}}$). In the last column are corresponding groups with hosting HSPs (also specified in Fig.~\ref{FSummary}).}
\begin{tabular}{|c|cc|ccc|p{13cm}|}
\hline
T&$|\ga|$&$L_{\ga}$&$f^{\tilde{\ga}}$&$|\tilde{\ga}|$&$L_{\tilde{\ga}}$&\dots, Group\,HSP$_1$\,HSP$_2$\dots,\dots\\\hline
$\alpha$& 1& 0& 1& 2& 2&
  $19SXY\Gamma$, $20X\Gamma$, $21S\Gamma$, $22S\Gamma$, $23SXY\Gamma$, $24X\Gamma$, $25S\Gamma$, $26S\Gamma$, $53SX\Gamma$, $54S\Gamma$, $55SX\Gamma$, $56S\Gamma$,
  $57SX\Gamma$, $58S\Gamma$, $59SX\Gamma$, $60S\Gamma$, $67\Gamma$, $68KL\Gamma$, $69\Gamma$, $70KL\Gamma$, $71K$, $72K$, $76KM\Gamma$, $77KM\Gamma$\\\hline
$\gamma$& 2& 0& 1& 4& 2&
  $62S$, $64S$\\\hline
$\zeta$& 2& 0& $\smatCD{1}{1}$&$\smatCD{2}{2}$&$\smatCD{2}{2}$&
  $39S$, $46S$, $53S\Gamma$, $54S\Gamma$, $55S\Gamma$, $56S\Gamma$, $57S\Gamma$, $58S\Gamma$,
  $59S\Gamma$, $60S\Gamma$\\\hline
$\eta$& 2& 0&${1\atop 1}$&${2\atop 2}$&${0\atop 2}$&
  $76\Gamma$, $77\Gamma$\\\hline
$\kappa$& 2& 1& 2& 2& 1&
  $7S$, $7Y$, $48Y$, $52X$\\\cline{2-7}
    & 2& 1& $\smatCD{1}{1}$&$\smatCD{2}{2}$&$\smatCD{1}{1}$&
  $38SY$, $39XY$, $41SY$, $42XY$, $43SY$, $45SY$, $46XY$, $62X$, $64X$\\\hline
$\lambda$& 2& 2& 2& 2& 2&
  $15SY$, $16SY$, $17XY$\\\hline
$\mu$& 2& 2&$\smatCD{1}{1}$&$\smatCD{2}{2}$&$\smatCD{0}{0}$&
  $40SY$, $43X$, $44XY$, $45X$, $63X$, $78\Gamma$, $79KL\Gamma$, $80K$\\\hline
$\nu$& 2& 2& $\smatCT{1}{1}{1}$&$\smatCT{1}{1}{2}$&$\smatCT{0}{0}{2}$&
  $67\Gamma$, $68KL\Gamma$, $69\Gamma$, $70KL\Gamma$, $71K$, $72K$, $76K$, $77K$\\\hline
$\xi$& 2& 2&$\smatCC{1}{1}{1}{1}$&$\smatCC{1}{1}{1}{1}$&$\smatCC{0}{0}{0}{0}$&
  $20SY$, $21XY$, $24SY$, $25XY$, $54X$, $56X$, $58X$, $60X$\\\hline
\end{tabular}
\end{table*}

However, not all completely linear band crossings remain such when spin space is added. Besides~\eqref{ESO}, this depends also on compatibility relation~\eqref{ECompat} between HSP and generic point stabilizer (co)IRs. Namely, the tensor product of the both sides of~\eqref{ECompat} by the spin representation $u$ can be found: obvious rule $u(G_{k_0}\downarrow G_{k_0+k})=u(G_{k_0+k})$ gives $(d^{(k_0\ga)}(G_{k_0})\otimes u(G_{k_0}))\downarrow G_{k_0+k}=\oplus_{i} f_i(d^{(k_0+k,\ga_i)}(G_{k_0+k})\otimes u(G_{k_0+k}))$. Then right and left sides are reduced in Clebsch-Gordan series.

As an illustration of mechanism how band splitting (the degeneracy of branches around a crossing point) is changed after the SO inclusion, let us consider the $u$-reducible case. The components $u_j$ ($j=1, 2$) are one-dimensional, and remain irreducible when subduced onto generic domain. Clearly, following the relation~\eqref{ESO}, each integer (orbital) (co)IR is decomposed onto two half-integer (co)IRs $d^{({k_0},\tilde{\ga}_j)}(G_{k_0})$ of the same dimension, equivalent to $d^{({k_0},\ga)}(G_{k_0})\otimes u_j(G_{k_0})$, giving essentially two independent energies. Applying further the compatibility relation leads to $(d^{(k_0\ga)}(G_{k_0})\otimes u_j(G_{k_0}))\downarrow G_{k_0+k}=\sum_i f^j_i(d^{k_0+k,\ga_i}(G_{k_0+k})\otimes u_j(G_{k_0+k}))$, which determines the degeneracy of branches around HSP for each group of bands counted by $j$ when SO is considered.

\squeezetable\begin{table*}\caption{\label{FSOGray} Influence of SO coupling to the type of the splitting with TR symmetry: each row denotes a particular type (label in the column T is used as superscript in Fig.~\ref{FSummary}) of transition from spinless case (described in the next two columns by degeneracy $|\ga|$ at the crossing point, linearity rank $L_{\ga}$, and Wigner's kind of subgroup IR $W_{\ga}$) to the spinfull case (in the following columns: frequency number $f^{\tilde{\ga}}$ in decomposition~\eqref{ESO}, crossing point degeneracy $|\tilde{\ga}|$, linearity rank $L_{\tilde{\ga}}$, and Wigner's kind of subgroup IR $W_{\tilde{\ga}}$). In the last column are corresponding groups with hosting HSPs (also specified in Fig.~\ref{FSummary}).}
\begin{tabular}{|c|ccc|cccc|p{11.5cm}|}
\hline
T&$|\ga|$&$L_{\ga}$ &$W_{\ga}$ &$f^{\tilde{\ga}}$&$|\tilde{\ga}|$&$L_{\tilde{\ga}}$ &$W_{\tilde{\ga}}$&\dots, Group\,HSP$_1$\,HSP$_2$\dots,\dots\\\hline
$\alpha$&1& 0& 1&$\smatCJ{1}$& $\smatCJ{2}$& $\smatCJ{2}$& $\smatCJ{-1}$&$1SXY\Gamma$, $10Y$, $13Y$, $65M$\\\hline
$\alpha$&1& 0& 1&$\smatCJ{1}$& $\smatCJ{2}$& $\smatCJ{2}$& $\smatCJ{0}$&$3SXY\Gamma$, $8SXY\Gamma$, $9X\Gamma$, $10S\Gamma$, $11SXY\Gamma$, $12X\Gamma$, $13S\Gamma$, $22Y$, $26Y$, $49SX\Gamma$, $50SX\Gamma$, $65\Gamma$, $67M$, $68M$, $69M$, $70M$, $73KM\Gamma$\\\hline
$\alpha$&1& 0& 1&$\smatCJ{1}$& $\smatCJ{2}$& $\smatCJ{2}$& $\smatCJ{1}$&$19SXY\Gamma$, $20X\Gamma$, $21\Gamma$, $22S\Gamma$, $23SXY\Gamma$, $24X\Gamma$, $25\Gamma$, $26S\Gamma$, $53SX\Gamma$, $54\Gamma$, $55SX\Gamma$, $56\Gamma$, $57SX\Gamma$, $58\Gamma$, $59SX\Gamma$, $60\Gamma$, $67\Gamma$, $68K\Gamma$, $69\Gamma$, $70K\Gamma$, $76KM\Gamma$, $77KM\Gamma$\\\hline
$\beta$&1& 0& 1&$\smatCJ{1}$& $\smatCJ{2}$& $\smatCJ{1}$& $\smatCJ{0}$&$4SXY\Gamma$, $5X\Gamma$, $35Y$, $74M$\\\hline
$\beta$&1& 0& 1&$\smatCJ{1}$& $\smatCJ{2}$& $\smatCJ{1}$& $\smatCJ{1}$&$27SXY\Gamma$, $28X\Gamma$, $29X\Gamma$, $30X\Gamma$, $31X\Gamma$, $32\Gamma$, $33\Gamma$, $34\Gamma$, $35S\Gamma$, $36S\Gamma$, $78M$, $79M$\\\hline
$\gamma$&2& 0& 0&$\smatCJ{1}$& $\smatCJ{4}$& $\smatCJ{2}$& $\smatCJ{-1}$&$21S$, $25S$\\\hline
$\gamma$&2& 0& 0&$\smatCJ{1}$& $\smatCJ{4}$& $\smatCJ{2}$& $\smatCJ{0}$&$54S$, $56S$, $58S$, $60S$\\\hline
$\gamma$&2& 0& 1&$\smatCJ{1}$& $\smatCJ{4}$& $\smatCJ{2}$& $\smatCJ{0}$&$39S$, $46S$, $52S$, $54S$, $56S$, $58S$, $60S$\\\hline
$\gamma$&2& 0& 1&$\smatCJ{1}$& $\smatCJ{4}$& $\smatCJ{2}$& $\smatCJ{1}$&$62S$, $64S$\\\hline
$\delta$&2& 1& 0&$\smatCJ{1}$& $\smatCJ{4}$& $\smatCJ{2}$& $\smatCJ{-1}$&$28SY$, $29SY$, $30SY$, $32X$, $33X$, $34X$\\\hline
$\delta$&2& 1& 1&$\smatCJ{1}$& $\smatCJ{4}$& $\smatCJ{2}$& $\smatCJ{-1}$&$7SY$, $15SY$, $16SY$, $17XY$, $48Y$, $52X$\\\hline
$\delta$&2& 1& 1&$\smatCJ{1}$& $\smatCJ{4}$& $\smatCJ{2}$& $\smatCJ{0}$&$38SY$, $39XY$, $41SY$, $42XY$, $43Y$, $45Y$, $46XY$, $62X$, $64X$\\\hline
$\epsilon$&2& 1& 1&$\smatCJ{1}$& $\smatCJ{4}$& $\smatCJ{1}$& $\smatCJ{0}$&$40SY$, $43X$, $44XY$, $45X$, $63X$\\\hline
$\zeta$&2& 0& 0&$\smatCD{1}{1}$& $\smatCD{2}{2}$& $\smatCD{2}{2}$& $\smatCD{0}{0}$&$49S\Gamma$, $50S\Gamma$\\\hline
$\zeta$&2& 0& 1&$\smatCD{1}{1}$& $\smatCD{2}{2}$& $\smatCD{2}{2}$& $\smatCD{1}{1}$&$53S\Gamma$, $54\Gamma$, $55S\Gamma$, $56\Gamma$, $57S\Gamma$, $58\Gamma$, $59S\Gamma$, $60\Gamma$\\\hline
$\eta$&2& 0& 0&$\smatCD{1}{1}$& $\smatCD{2}{2}$& $\smatCD{0}{2}$& $\smatCD{-1}{0}$&$65\Gamma$\\\hline
$\eta$&2& 0& 0&$\smatCD{1}{1}$& $\smatCD{2}{2}$& $\smatCD{0}{2}$& $\smatCD{0}{0}$&$73\Gamma$\\\hline
$\eta$&2& 0& 1&$\smatCD{1}{1}$& $\smatCD{2}{2}$& $\smatCD{0}{2}$& $\smatCD{0}{1}$&$67\Gamma$, $68\Gamma$, $69\Gamma$, $70\Gamma$\\\hline
$\eta$&2& 0& 1&$\smatCD{1}{1}$& $\smatCD{2}{2}$& $\smatCD{0}{2}$& $\smatCD{1}{1}$&$76\Gamma$, $77\Gamma$\\\hline
$\theta$&2& 0& 1&$\smatCD{1}{1}$& $\smatCD{2}{2}$& $\smatCD{1}{1}$& $\smatCD{0}{0}$&$32S$, $34S$\\\hline
$\iota$&2& 1& 0&$\smatCD{1}{1}$& $\smatCD{2}{2}$& $\smatCD{2}{2}$& $\smatCD{-1}{-1}$&$5SY$, $36Y$\\\hline
$\iota$&2& 1& 1&$\smatCD{1}{1}$& $\smatCD{2}{2}$& $\smatCD{2}{2}$& $\smatCD{0}{0}$&$31SY$, $32Y$, $33Y$, $34Y$\\\hline
$\kappa$&2& 1& 0&$\smatCD{1}{1}$& $\smatCD{2}{2}$& $\smatCD{1}{1}$& $\smatCD{-1}{-1}$&$9SY$, $12SY$\\\hline
$\kappa$&2& 1& 1&$\smatCD{1}{1}$& $\smatCD{2}{2}$& $\smatCD{1}{1}$& $\smatCD{0}{0}$&$20SY$, $21XY$, $24SY$, $25XY$, $54X$, $56X$, $58X$, $60X$\\\hline
$\mu$&2& 2& 0&$\smatCD{1}{1}$& $\smatCD{2}{2}$& $\smatCD{0}{0}$& $\smatCD{-1}{0}$&$66K$\\\hline
$\mu$&2& 2& 0&$\smatCD{1}{1}$& $\smatCD{2}{2}$& $\smatCD{0}{0}$& $\smatCD{0}{0}$&$75K$\\\hline
$\mu$&2& 2& 1&$\smatCD{1}{1}$& $\smatCD{2}{2}$& $\smatCD{0}{0}$& $\smatCD{0}{1}$&$71K$, $72K$\\\hline
$\mu$&2& 2& 1&$\smatCD{1}{1}$& $\smatCD{2}{2}$& $\smatCD{0}{0}$& $\smatCD{1}{1}$&$79K$, $80K$\\\hline
$\nu$&2& 2& 0&$\smatCD{1}{2}$& $\smatCD{2}{1}$& $\smatCD{2}{0}$& $\smatCD{0}{1}$&$73K$\\\hline
$\nu$&2& 2& 1&$\smatCT{1}{1}{1}$& $\smatCT{1}{1}{2}$& $\smatCT{0}{0}{2}$& $\smatCT{1}{1}{1}$&$68K$, $70K$, $76K$, $77K$\\\hline
$o$&4& 2& 0&$\smatCD{1}{1}$& $\smatCD{4}{4}$& $\smatCD{1}{1}$& $\smatCD{-1}{-1}$&$43S$, $45S$\\\hline
$\pi$&4& 2& -1&$\smatCC{1}{1}{1}{1}$& $\smatCC{2}{2}{2}{2}$& $\smatCC{1}{1}{1}{1}$& $\smatCC{-1}{-1}{-1}{-1}$&$33S$\\\hline
\end{tabular}
\end{table*}

We calculated the decompositions~\eqref{ESO} for the both cases without and with TR symmetry. Results with crossing bands are presented in the Table~\ref{FSOOrdinary} for ordinary groups and~\ref{FSOGray} for gray groups, together with linearity rank. Extracting the data from these tables, i.e. analysing~\eqref{ESO} for all possible dimensions (1, 2, and 4) of (co)IRs, different ways how SO may affect band crossings are listed bellow, where notation $|\ga| \SO \oplus_{\tilde{\ga}} |\tilde{\ga}|$ is used to explicate the dimensions of the allowed representations in spinless and spinful cases. Non-crossing cases correspond to linearity rank 0.\\
\begin{enumerate}
  \item[$\bullet$] $1 \SO 2$. SO induces transition from an orbital nondegenerate band (no crossing) to a band crossing, with one of the following dispersions:\\
    \phantom{X}($\alpha$)\label{SO12D0t2} 1DC;\\
    \phantom{X}($\beta$)\label{SO12D0t1} linearity rank 1.
  \item[$\bullet$] $2 \SO 4$. Transitions from 2D integer (co)IR are:\\
    \phantom{X}($\gamma$)\label{SO24D0t2} a 2-fold orbital band (no crossing) becomes 4-degenerate point with (modified) PF or 2DC; \\
    \phantom{X}($\delta$)\label{SO24D1t2} 2D crossing point of linearity rank 1 yields 4-degenerate band crossing with PF, FT or 2DC;\\
    \phantom{X}($\epsilon$)\label{SO24D1t1} 2D crossing point of linearity rank 1 becomes 4D crossing with linearity rank 1.
  \item[$\bullet$] $2 \SO 2 \oplus 2$. When 2D integer (co)IR produces two 2D half-integer (co)IRs, possible patterns are:\\
    \phantom{X}($\zeta$)\label{SO222D0t22} single 2-fold orbital band (no crossing) yields two 1DC (differing in energy);\\
    \phantom{X}($\eta$)\label{SO222D0t200} single 2-fold orbital band (no crossing) becomes a 1DC and a 2-fold band (without crossing);\\
    \phantom{X}($\theta$)\label{SO222D0t11} single 2-fold orbital band (no crossing) gives two 2-degenerate crossings of linearity rank 1;\\
    \phantom{X}($\iota$)\label{SO222D1t22} 2-degenerate point of linearity rank 1 gives two 1DC;\\
    \phantom{X}($\kappa$)\label{SO222D1t11} 2-degenerate point of linearity rank 1 gives two 2-degenerate linearity rank 1 crossings;\\
    \phantom{X}($\lambda$)\label{SO222D2t22} spinless 1DC yields two 1DC;\\
    \phantom{X}($\mu$)\label{SO222D2t00} spinless 1DC transforms into two 2-fold band (gap opening pattern).
  \item[$\bullet$] $2\SO 2 \oplus 1 \oplus 1$. Transition from a spinless 1DC crossing to:\\
    \phantom{X}($\nu$)\label{SO2211D2t211} 1DC and two non-degenerate bands (cone preserving).
  \item[$\bullet$] $2\SO 1 \oplus 1 \oplus 1 \oplus 1$: Another gap opening pattern, where a spinless 1DC splits into\\
    \phantom{X}($\xi$)\label{SO21111D2t0} four non-degenerate bands (no crossing).
  \item[$\bullet$] $4 \SO 4 \oplus 4$: One way to split spinless FT dispersion (4D allowed integer representation) is to\\
    \phantom{X}($o$)\label{SO4to44} two 4-fold crossings of the linearity rank 1.
  \item[$\bullet$] $4\SO 2\oplus2\oplus2 \oplus 2$: Also, spinless FT dispersion may be transformed into  \\
    \phantom{X}($\pi$)\label{SO4to2222} four 2D crossings of linearity rank 1.
\end{enumerate}

\begin{figure*}
\includegraphics[height=0.85\textheight,width=0.85\textwidth]{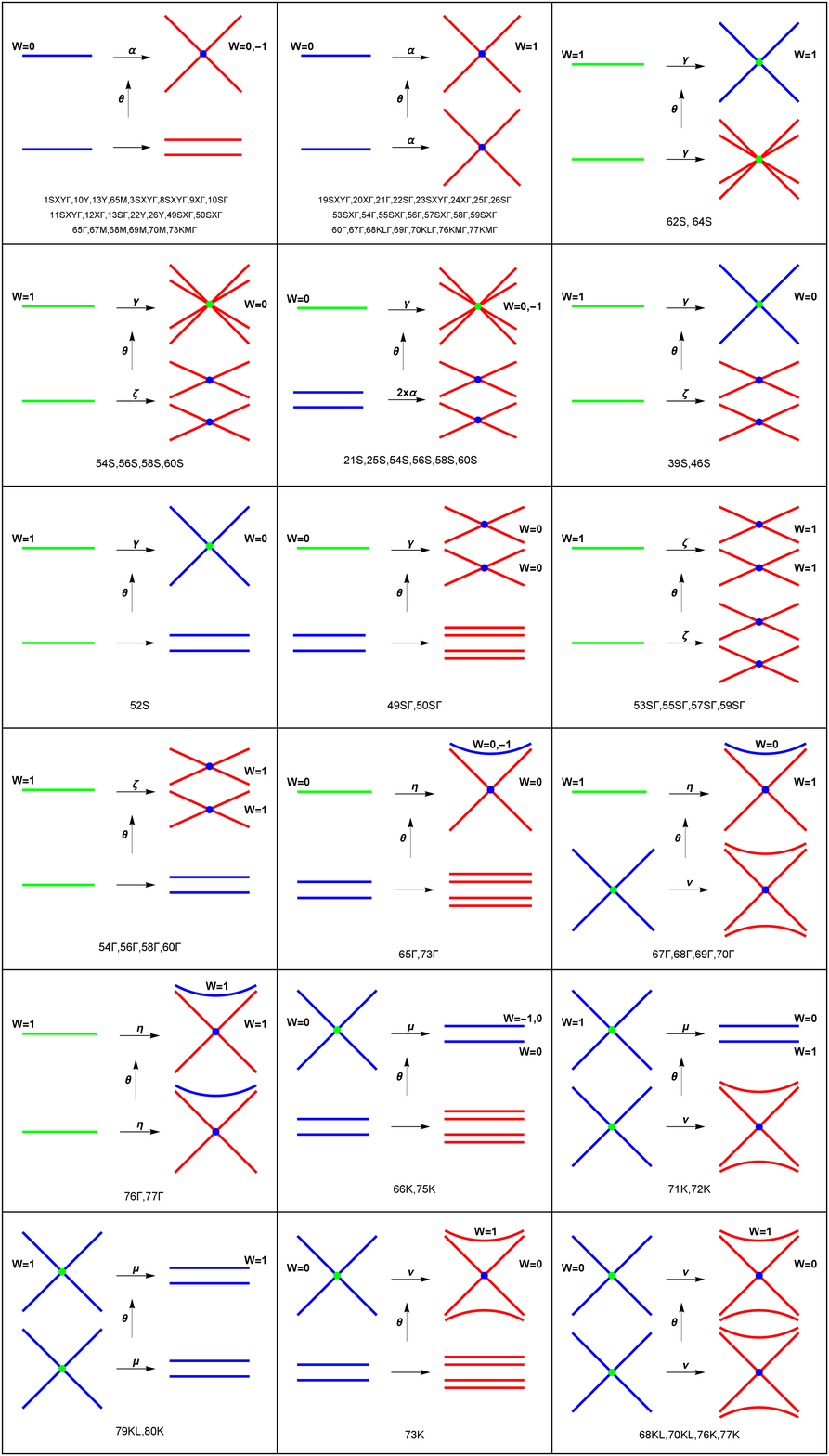}
\caption{\label{FSOTR} Manifestations of SO coupling and TR symmetry. Each figure describes transitions within a cluster at HSP (cluster and HSP are indicated at the bottom): horizontal arrows are for transitions from single to double groups (with the Greek letter indicating type from Tables~\ref{FSOOrdinary} and~\ref{FSOGray}), while vertical ones are from ordinary to gray (with indicated Wigner kind of hosting coIR from Tab.~\ref{FSOGray}). Color of the bands is degeneracy in orbital-spin space: red, blue and green are for degeneracy 1, 2 and 4.}
\end{figure*}

\subsection{Time-reversal symmetry}

The role of TR symmetry is clarified through the transition from ordinary $L$ to gray groups $G$. This involves magnetic (black-and-white, as well) little groups, and possibly new strata (with change of the irreducible domain), including HSPs of $G$ not characterizing the corresponding $L$. An enlarged stabilizer of a momentum $k$ may give rise to an enlarged degeneracy of the energy in $k$, while enlarged star necessarily enlarges the dimension of the associated coIR. In fact, the impact of TR symmetry is essentially encoded in the algorithm for co-representations construction. Irreducible co-representations~\cite{DIMMOCK62,BRADLEY68} of $G_k$ are derived from IRs of $L_k$: each real IR (Wigner's I kind) of $L_k$ is extended to co-IR of $G_k$, a quaternion IR (II kind, equivalent to its conjugate, but without equivalent real IR) gives co-IR of the double dimension, while two mutually conjugate complex IRs give one co-IR of the double dimension. Hence, besides the case of an ordinary stabilizer $G_k=L_k$, TR symmetry preserves the HSP degeneracy also for crossings hosted by HSP invariant under magnetic group, but with allowed coIR determined by a real subgroup IR. On the other hand, the HSP degeneracy may be doubled for magnetic stabilizers with quaternion or complex subgroup IR. However, even when HSP degeneracy remains the same, the dispersion need not stay completely linear, and its shape may be not preserved. The enlarged group by TR imposes new conditions on Hamiltonian parameters and also affect the compatibility relations.

For this purpose to each of the stabilizer (co)IR we assign the number $W$, which shows whether it is composed of two (mutually non-equivalent $W=0$ or equivalent $W=-1$) or one ($W=1$) subgroup IR. It is given as the last entry in the Tables~\ref{FSOOrdinary} and~\ref{FSOGray} to enable tracking the role of TR symmetry. To illustrate, let us consider, for example, the transition $1 \SO 2$ ($\alpha$) to 1DC described in the Subsec.~\ref{SSSO}. In the Table~\ref{FSOGray} this appears in 3 rows mutually differing by the last entry (column $W_{\ti{\alpha}}$). In the third case, when both integer and half-integer coIRs carry the value $W_{{\alpha}}=W_{\ti{\alpha}}=1$, the corresponding groups appear also in Tab.~\ref{FSOOrdinary}; this means that this type of transition is preserved under TR symmetry. On the contrary, the remaining two cases (with last entries $0$ and $-1$ for half-integer coIRs) do not appear in Tab.~\ref{FSOOrdinary}. This is expected since herein a conical dispersion in gray DLG is hosted by the half-integer coIR composed of two 1D subgroup half-integer IRs. Thus, breaking TR symmetry in these cases leads to non-crossing bands. The both situations are sketched in Fig.~\ref{FSOTR}.

One can further similarly analyse relations between ordinary and gray groups case-by-case. In this way, combining the results from the both Tables ~\ref{FSOGray} and ~\ref{FSOOrdinary}, different cluster processes can be found. Some of them are illustrated in Fig.~\ref{FSOTR}; the skipped cases are with linearity rank 1 either in spinless or in spinfull case.

\section{Conclusion}

The linear dispersions at high symmetry points and underlying effective models allowed by integer and half-integer 2D and 4D (co)IRs are studied. Different dispersion types linear in all directions are classified and listed, completing thus the results existing in literature.
Having these data at disposal, it was possible to analyze influence of SO coupling and time reversal symmetry to interrelate dispersions within the same cluster of the single/double ordinary/gray layer groups.

Summarizing results, firstly note that the LG clusters 2, 6, 14, 18, 37, 47, 51, 61, all of them being centrosymmetric, do not support linear band crossing in HSPs at al, while 4, 27, 35 and 74 do not support fully linear (with linearity rank 2), but have linearity rank 1 band crossings (see Tab.~\ref{FSOGray}). Further, as visible in Table~\ref{FSummary}, the only fully linear 2D band crossing model in HSPs is 1DC. Notably, these are hosted at TRIM and non-TRIM points in ordinary single, as well as in ordinary and gray double groups in both symmorphic and non-symmorphic cases. In the remaining (gray single) groups, 1DC occurs only in $K$ (thus not TRIM) point of some (symmorphic) groups~\cite{Ja16,Ja16Ad}.

As for 4D models, inclusion of spin gives four-fold degenerate point with PF in two double groups (LG 62 and LG 64), while TR gives rise to FT dispersion~\cite{Ja17} in 3 gray LGs. The presence of both spin and TR give rise to 4D coIRs in 26 gray double layer groups. Only 3 of them (7,48,52) are without  special lines; their special points are surrounded by generic points with 2D allowed coIRs, enabling only 2DC dispersions.  In all other 4D cases, besides 2DC cases (for 2D generic allowed coIRs), nondegenerate generic coIRs enable also 4-band dispersion structures, but special lines with degenerate coIRs impose touching of pairs of bands, restricting linear rank 2 dispersions to PF and FT types. PF and FT types appear in noncentrosymmetric gray DLGs with a non-symmorphic symmetry: FT in 2 groups, and PF in 10 groups in total~\cite{DamljanovicJPC20}. Degeneracy of the generic allowed representations in centrosymmetric gray DLGs admits 2DC dispersions, as it was proposed~\cite{JaKe15}; actually, this is realized in 15 of these groups, as in the remaining 3 (40, 44 and 63, nonsymmorphic) the dispersion is linear along a single direction, while the second one is special line (at BZ edge) with single 4D allowed coIR, thus becoming four-fold degenerate nodal line. In particular, concerning IDs, 2DC is found in three HSPs $X, Y, S$ (gray DLGs 39, 46), in two HSPs $X, S$ (52, 62, 64), in two HSPs $Y, S$ (7, 15, 16, 38, 41), in two HSPs $X, Y$ (17, 42), and single point $Y$ (gray DLGs 43, 45, 48). In the groups 43 and 45 additional 4-fold band crossings, as required by fermion doubling theorem~\cite{WiBr18}, are at $X$ and $S$, but have linearity rank 1 (Tab.~\ref{FSOGray}). Concerning the whole BZ, note that for the groups 48, 52, 62, and 64 points $X$ and $Y$ are symmetry related. Thus, for engineering Dirac semimetals, it is particularly important to single out group 48, since effectively one need to tune band contacts only at a single point, \emph{i.e.} for filling $4n+2$, both (symmetry related) cones in BZ are on the Fermi level, if there are no additional electron or hole pockets.

It is interesting that simultaneously 2D and 4D completely linear dispersions are hosted only by the gray DLGs 21, 25, 32, 33, 34, 54, 56, 58, 60 (note that in these groups there are also HSPs with linearity rank 1).

Inclusion of the spin orbit interaction causes various effects on the HSPs' dispersions, including gap closing ($\alpha$, $\gamma$, $\eta$), gap opening ($\nu$), cone preserving ($\xi$), cone splitting ($\lambda$) scenarios (discussion about the cases with the linearity rank 1 is skipped). For example, an isotropic 1DC in gray LGs~\cite{Ja16,Ja16Ad}, which is preserved ($\xi$) by SO perturbation also in gray DLG, is at $K$ point in symmorphic cluster 68, 70, 73, 76, 77. Similar analysis reported in Ref.~\cite{LuoPRB20} omitted symmorphic gray DLG 73. Concerning the TR symmetry breaking, we found also that the cone persists at $K$ point in corresponding LGs and DLGs 68, 70, 76, 77, except in the group LG and DLG 73, where the vanishing TR symmetry opens a gap.

Besides spinless to spinfull transition, we examined influence of TR symmetry to dispersion at crossing point. Addition of TR symmetry may preserve or double the degeneracy in HSP. Concerning the preserved double degeneracy, our results single out the cases where 1DC appears both with and without TR symmetry, as well as those when TR even prevent linearity of dispersion. On the other hand, TR symmetry in centrosymmetric groups 62 and 64, although does not change 4-fold degeneracy, modifies the dispersion type: in ordinary double groups two generic nondegenerate allowed IRs enable two positive (and two negative) bands touching along special lines (with single degenerate allowed IR); TR symmetry joins these IRs in a single 2D allowed coIR, transforming PF to 2DC dispersion.

Focusing on TR symmetric materials without and with SO from the literature, we further discuss applicability of our results. The frequently elaborated honeycomb lattice belongs to LG 80 with $K$ point hosting Dirac cone being gapped by SO. That is symmetry prediction confirmed by DFT calculations in honeycomb lattices of $\mathrm{C}$, $\mathrm{Si}$, $\mathrm{Ge}$, $\mathrm{Sn}$ or $\mathrm{Pb}$ elements~\cite{VVDDYak17, VVDDGutSch17}. Buckled honeycomb lattice belongs to LG 72 with the same behaviour of bands near $K$ as in LG 80. Tight binding model on $\mathrm{Si}, \mathrm{Ge}$ and $\mathrm{Se}$ elemental lattices~\cite{VVDDLiHu11} and DFT band structure of $\mathrm{As_2X_2}$ ($\mathrm{X}=\mathrm{Cl}, \mathrm{F}, \mathrm{I}, \mathrm{Br}$) monolayers~\cite{VVDDTaSu16} confirm our predictions. Similarly, Dirac cones split by SO near $K$ point shows LG 66 with nonmagnetic high buckled $\mathrm{Co_2C_{18}H_{12}}$ as DFT-example~\cite{VVDDMaDa14}. On the other hand LG 77 supports Dirac cones at $K$ both without and with SO, with monolayer $\mathrm{FeB_2}$~\cite{VVDDZhLi16} and $\mathrm{HfB_2}$~\cite{VVDDLiWa19} as DFT-examples. Square LG 64 supports Dirac cones at $X$ and $S$ only in the presence of SO interaction; this is confirmed by DFT band structure of $\mathrm{MX}$ compounds ($\mathrm{M=Sc, Y; X=S, Se, Te}$)~\cite{VVDDGuZh20} as well as  in $X$ point ($S$ point was not discussed since the corresponding energies are too far from the Fermi level) in ARPES experiments and DFT calculations in synthesised layered 3D $\mathrm{ZrSiS}$~\cite{VVDDScAl16} and numerically in monolayer $\mathrm{HfGeTe}$~\cite{VVDDGuLi17}. Experimentally synthesised $\ga$-Bismuthene belongs to LG 42 and hosts spinfull Dirac cones at $X$ and $Y$ points, as confirmed by micro-ARPES technique and DFT calculations~\cite{VVDDKoBr20}.

Among already reported structures with PF or FT dispersions are monolayer $\mathrm{GaXY}$ ($\mathrm{X=Se, Te; Y=Cl, Br, I}$), with non-centrosymmetric symmetry LG 32 providing SO caused Dirac cones at $X$ point and PF at $Y$ point. Indeed, fourfold degeneracy at $Y$ point (called Dirac point in~\cite{VVDDWuJi19}) splits linearly away from it, as justified numerically~\cite{VVDDWuJi19} (dispersion near $X$ point was not discussed more closely). DFT band structure of monolayer Ta3SiTe6 and Nb3SiTe6~\cite{VVDDLiLi18} requires particular attention. Corresponding structure with space group $Pmc2_1$ (SG 26 in notation~\cite{VVDDITCA}) is obtained by periodic distribution of monolayers along vertical axis. The monolayers may be of the symmetry either LG 28 or LG 29; these two groups are similar, both with the  horizontal screw axis of order two, and two planes, the vertical one is mirror and the horizontal glide in LG 28, while in LG 29  the vertical is glide and the horizontal is mirror.
LG 28 should host PF dispersions at the points $Y$ and $S$, with low energy effective six-parameters Hamiltonian. However, monolayers $\mathrm{Ta_3SiTe_6}$ and $\mathrm{Nb_3SiTe_6}$ have horizontal symmorphic mirror plane~\cite{VVDDLiLi18}, and their symmetry group is LG 29,  with FT dispersions (special case of PF) at $Y$ and $S$ points, and effective Hamiltonian having four independent real parameters. Indeed, linear dispersion in $Y$ and $S$ points are reported~\cite{VVDDLiLi18} (instead of minimal 4 parameters authors use 6 as for LG 28, which can not affect the result).

Since surfaces of (semi-infinite) 3D single crystals are also periodic in two directions, some layer groups are also wallpaper groups being the symmetries of surfaces. Those contain symmetry elements that do not flip the surface: perpendicular rotational axes of order two, three, four, or six, and perpendicular mirror, or glide planes. It may happen that surface reconstruction or adding atoms at surface in regular manner can lower the symmetry. Such is the case for (110) surface of silicon, where FT dispersion was found experimentally~\cite{VVDDZdyb20}. FT dispersion was caused by the Coulomb interaction (described by gray LGs) rather than by the relativistic corrections (described by grey DLGs) so linear dispersion is maintained over wide energy range. In addition, BZ of reconstructed surface shrinks, so that another FT dispersion at the centre of rectangular surface BZ is obtained by intersection from FT bands originating from the corners. This might explain why FT dispersion at $\ol{X}$ of Si(110) surface, seen in ARPES~\cite{VVDDZdyb20}, remained intact by different surface reconstruction types.

3D TIs are known~\cite{HasanKane} to have large SO coupling that causes Dirac cones at surface states. Our results apply also to TIs with the remark that only surface states that fall within the bulk gap are investigated in the literature, since they give rise to surface conductivity. The surface states with the energy within the bulk gap, are identified by analysis of topological properties of bulk bands (via bulk-boundary correspondence) and cannot be predicted by group theory alone. 3D compounds $\mathrm{Bi_2Se_3}$, $\mathrm{Bi_2Te_3}$, $\mathrm{Sb_2Te_3}$ and $\mathrm{Sb_2Se_3}$ belong to the SG 166 ($R\bar{3}m$) with (111) surface with symmetry gray DLG 69 so Dirac cones are expected in $\ol{\Gamma}$ and $\ol{M}$ of the surface BZ. DFT calculations show that first three materials have surface Dirac cone at $\ol{\Gamma}$ within the bulk gap, while states near $\ol{M}$ fall far out of the bulk gap and were not shown. On the other hand the last compound $\mathrm{Sb_2Se_3}$ does not have surface states in the gap and it is not TI~\cite{VVDDZhLi09}. Surface low energy effective Hamiltonian near $\ol{\Gamma}$ has one real parameter, in accordance with our results. Surface Dirac cone in $\ol{\Gamma}$  has been seen in ARPES experiments in $\mathrm{Bi_2Te_3}$ and $\mathrm{Sb_2Te_3}$~\cite{VVDDHsXi09}. Similarly, 3D compound $\mathrm{LaBi}$ crystalises in SG 225 ($Fm\bar{3}m$) with (001) surface having symmetry LG 55. SOC Dirac cones are expected to appear on $\ol{S}$, $\ol{\Gamma}$, and $\ol{X}$ points of the BZ. ARPES experiments supported by DFT calculations show Dirac cones at $\ol{\Gamma}$ and $\ol{S}$ in the bulk gap, while bands near $\ol{X}$ were outside the gap~\cite{VVDDNaWu17}. Theoretically proposed 3D compound $\mathrm{Sr_2Pb_3}$, that belongs to SG 127 ($P4/mbm$) and its (001) surface to LG 56 (wallpaper group 12 in notation~\cite{VVDDITCA}), is expected to be non-symmorphic TI~\cite{WiBr18}. Our result show that SO causes Dirac cone at $\ol{\Gamma}$ and PF at $\ol{M}$ point for LG 56. DFT band structure show linear dispersions from fourfold degenerate energy at $\ol{M}$~\cite{WiBr18}. Their effective low energy Hamiltonian has two independent real parameters and suggests that the dispersion is Dirac-like (2DC in our notation). Necessary splitting that causes bands along $\ol{M}-\ol{\Gamma}$ to be non-degenerate (as required by symmetry) was attributed to quadratic corrections to the effective Hamiltonian~\cite{WiBr18}. Our analysis indicates that the dispersion at $\ol{M}$ should be PF, with three-parameters Hamiltonian and with bands along $\ol{M}-\ol{\Gamma}$ being non-degenerate already in the linear approximation.

The presented theoretical framework is straightforwardly extendable to (ferro/anti-ferro) magnetic systems invariant under black-and-white ordinary or double groups. Also, it can be used on an equal footing to analyse higher order dispersion terms, dispersions in the vicinity of special lines which occur in 2D BZ of layer materials, as well as to clarify the cases with single linear direction in energy.



\begin{acknowledgments}
Authors acknowledge funding by the Ministry of Education, Science and Technological Development of the Republic of Serbia provided by the Institute of Physics (V.D.), Faculty of Physics (N.L. and M.D.) and Serbian Academy of Sciences and Arts (M.D.).
\end{acknowledgments}


%


\end{document}